\documentclass[english,aps,prb,twocolumn,showpacs,superscriptaddress,nofootinbibfloatfix]{revtex4-1}
\usepackage[T1]{fontenc}
\usepackage[latin1]{inputenc}
\usepackage{graphicx}
\usepackage{bm}
\usepackage{amssymb}
\usepackage{color}
\usepackage[usenames,dvipsnames]{xcolor}
\usepackage{amsmath}
\usepackage{amstext}
\usepackage{latexsym}
\usepackage[colorlinks=true,citecolor=Cerulean,linkcolor=RubineRed,urlcolor=Cerulean]{hyperref}

\usepackage{mathptmx} 
\usepackage{verbatim}

\usepackage{amsfonts}
\usepackage{epsfig}
\usepackage{dsfont}
\usepackage{subfigure}
\usepackage{mathrsfs}
\usepackage{arydshln,leftidx,mathtools}

\begin{document}




\pdfoutput=1

\title{Truncated Calogero-Sutherland models}


\author{S. M. Pittman}
\affiliation{Department of Physics, Harvard University, 
Cambridge, MA 02138, USA}
\author{M. Beau}
\affiliation{Department of Physics, University of Massachusetts, Boston, MA 02125, USA}
\author{M. Olshanii}
\affiliation{Department of Physics, University of Massachusetts, Boston, MA 02125, USA}
\author{A. del Campo}
\affiliation{Department of Physics, University of Massachusetts, Boston, MA 02125, USA}


\begin{abstract}
A one-dimensional  quantum many-body system consisting of  particles  confined in a harmonic potential and subject to   finite-range two-body  and  three-body inverse-square interactions is introduced.  The range of the interactions is set by truncation beyond a number of neighbors  and can be tuned to interpolate between the Calogero-Sutherland model and a system with nearest and next-nearest neighbors interactions discussed by Jain and Khare.
The model also includes the Tonks-Girardeau gas describing impenetrable bosons as well as a novel extension with truncated interactions.
While the ground state wavefunction takes a truncated Bijl-Jastrow form, collective modes of the system are found in terms of multivariable symmetric polynomials. 
We numerically compute the density profile, one-body reduced density matrix, and momentum distribution  of the ground state as a function of the range $r$ and the interaction strength. 
\end{abstract}






\maketitle

\def\q{{\bf q}}

\def\G{\Gamma}
\def\L{\Lambda}
\def\la{\lambda}
\def\g{\gamma}
\def\al{\alpha}
\def\s{\sigma}
\def\e{\epsilon}
\def\k{\kappa}
\def\ve{\varepsilon}
\def\l{\left}
\def\r{\right}
\def\te{\mbox{e}}
\def\d{{\rm d}}
\def\t{{\rm t}}
\def\K{{\rm K}}
\def\N{{\rm N}}
\def\H{{\rm H}}
\def\la{\langle}
\def\ra{\rangle}
\def\om{\omega}
\def\Om{\Omega}
\def\vep{\varepsilon}
\def\wh{\widehat}
\def\tr{\rm{Tr}}
\def\da{\dagger}
\def\iz{\left}
\def\zi{\right}
\newcommand{\beq}{\begin{equation}}
\newcommand{\eeq}{\end{equation}}
\newcommand{\beqa}{\begin{eqnarray}}
\newcommand{\eeqa}{\end{eqnarray}}
\newcommand{\intf}{\int_{-\infty}^\infty}
\newcommand{\into}{\int_0^\infty}



Quantum systems with inverse-square interactions play a prominent role across a wide variety of fields.
They are ubiquitous in many-body physics  where they have facilitated the understanding of fractional quantum Hall effect
and generalized exclusion statistics \cite{Haldane91,Wu94,MS94}.
Historically, their study played a key role in understanding the integrability of systems with long-range interactions and the development of asymptotic Bethe ansatz \cite{Sutherland71,Sogo93,Sutherland95}. 
Following the pioneering works by Dyson \cite{dyson1962statistical} and Sutherland \cite{Sutherland71},  their connection to random matrix theory has remained a fruitful line of research \cite{Mehta00,Forrester10}. 
They have also found applications in blackhole physics \cite{GT99} and  conformal field theory \cite{CSCFT,Cardy04,Vasiliev95,Marotta96,Cadoni04,Estienne12}.
More recently, they have been explored in the context of quantum decay of many-particle unstable systems \cite{delcampo16} 
and  in the study of thermal machines in quantum thermodynamics \cite{JBD15,BJD16}.

In the one-dimensional continuum space, a many-body system with inverse-square interactions is generally known as the Calogero-Sutherland model (CSM) \cite{Calogero71,Sutherland71,Polychronakos06}. The CSM occupies a privileged status among exactly-solvable models as a source of inspiration \cite{Ha96,Sutherland04,KK14}. In its original form, it describes one-dimensional bosons with inverse-square interactions of strength $\lambda$ that exhibit a universal Luttinger liquid behavior \cite{CSCFT}. 
Under harmonic confinement, this interacting Bose gas is equivalent to an ideal gas of particles with generalized exclusion statistics \cite{MS94}. It is then referred to as the rational Calogero gas. Its connection with random matrix theory is manifested in the ground-state probability density distribution, which takes the form of the joint probability density for the eigenvalues of the Gaussian $\beta$-ensemble with Dyson index $\beta=2\lambda$ \cite{Sutherland71,Forrester10}. While the CSM has shed new light on  interacting systems, 
it  can be mapped to a set of noninteracting harmonic oscillators \cite{Kawakami93,VOK94}.
Further, the CSM can be  extended to  account for fermionic statistics, internal degrees of freedom, and additional interactions, e.g., of Coulomb type \cite{Kawakami93,VOK94}. 
In particular, when the range of the interaction is truncated to nearest-neighbors, the system is quasi-exactly solvable if supplemented with a three-body term, as discussed by Jain and Khare \cite{JK99}.

 In this work, we introduce a family of one-dimensional quantum systems with tunable inverse-square interactions 
 that extend over a finite number of neighbors. These systems generally involve  pair-wise interactions, that can be either repulsive or attractive, as well as  three-body attractive interactions.   The CSM and the Jain-Khare model are recovered as particular limits.   Further instances within this family include the Tonks-Girardeau gas, describing impenetrable  bosons in one-dimension \cite{Girardeau60,GWT01}, and its generalization to finite-range interactions with two- and three-body terms.  We shall refer to all these systems as  truncated Calogero-Sutherland models (TCSM). They constitute a rare instance among solvable models as they involve interactions that can be tuned both as a function of the strength and range. 
The TCSM can also be understood as a CSM with screening, 
where the inverse-square field becomes suppressed when it attempts 
to penetrate more than a certain threshold number of particles.
The TCSM is the first, to our knowledge, quasi-solvable model 
that supports screening of external or interparticle interaction
forces by other particles, an effect otherwise 
present in a range of physical settings, 
from laser cooling \cite{dalibard1988_203,fioretti1998_415} to solid state \cite{kittel1953_book}.    
  The ground state of the TCSM is shown to be described by a truncated Bijl-Jastrow form.  Further, we find collective excitations in terms of multivariable symmetric polynomials and determine the  level degeneracy of this particular set of states.  We numerically compute the ground state density profile and the one-body reduced density matrix and discuss the role of the strength and finite range of the interactions in both  local and nonlocal one-body correlation functions.


\section{Model and ground-state properties}
The ground state of the CSM \cite{Calogero71,Sutherland71} is well known to be exactly described by the product of single- and two-particle correlations, i.e.,  a Bijl-Jastrow form \cite{Calogero71,Sutherland71,Kawakami93,VOK94}. The same holds true for the model  with  inverse-square interactions between nearest neighbors  and an attractive three-body term
initially discussed by Jain and Khare \cite{JK99,ASK01,BMK01,Ezung05}.
This observation prompts us to consider a ground state described by the many-body wavefunction 
\beqa
\label{gsansatz}
\Psi_0({\bf x})=C_{\N,\lambda, r}^{-1/2}\phi({\bf x})\varphi({\bf x}),
\eeqa
with
\beqa
\label{gaussian}
\phi({\bf x})=\exp\left[-\frac{m\om}{2\hbar}\sum_{i=1}^{\N}x_i^2\right],
\eeqa
and
\beqa
\label{zeqn}
\varphi({\bf x})=
\prod_
{\substack{i<j \\
|i-j|\leq r}}
(x_i-x_j)^{\lambda}.
\eeqa
Here, ${\bf x}=(x_1,\dots,x_\N)\in \mathbb{R}^\N$.
We shall refer to the corresponding probability distribution as the finite-range Dyson model 
\beqa
\Psi_0^2=C_{\N,\lambda, r}^{-1}\exp\left[-\frac{m\om}{\hbar}\sum_{i=1}^{\N}x_i^2\right]
\prod_
{\substack{i<j \\
|i-j|\leq r}}
(x_i-x_j)^{2\lambda},
\eeqa
as it reduces  for $r=1$ to the short-range Dyson model \cite{JK99} and for $r=\N-1$ to the Gaussian $\beta=2\lambda$ ensemble in random matrix theory for $\lambda=1/2,1,2$  \cite{Mehta00,Forrester10}. 

 For $r< \N-1$, the wavefunction does not have full permutation symmetry upon exchange of any two coordinates.  We therefore work within an ordered sector, $\mathcal{R}=\{\mathbf{x}\in \mathbb{R}^\N\ | x_1 > x_2 > \cdots >  x_\N \}$.  
In this sector, the normalization constant is given by the multi-dimensional integral $C_{\N,\lambda, r}=\int_{\mathcal{R}} d^{\N}{\bf x}\, \Psi_0^2$.
While  the full permutation symmetry of the system is broken upon truncation of the interaction range, it can be restored by symmetrizing the Hamiltonian over all permutations of coordinates, as discussed in Ref.~\citenum{ASK01} for the Jain-Khare model. The TCSM then represents a quantum fluid as opposed to a quantum chain of impenetrable particles. 
For instance, the corresponding ground state is obtained by
\beqa
\varphi(\mathbf{x})\rightarrow \frac{1}{\sqrt{\N!}}\sum_{P\in S_\N}\prod_
{\substack{P(i)<P(j) \\
|P(i)-P(j)|\leq r}}
(x_{P(i)}-x_{P(j)})^{\lambda}\,\Theta_{P}(\mathbf{x}), \label{symmetric}
\eeqa
where $P$ runs over the symmetric group  $S_\N$ with $\N!$ permutations and the ordering of the sector is set by $\Theta_{P}(\mathbf{x})=\{\mathbf{x}\in \mathbb{R}^\N|\ x_{P(1)}> x_{P(2)}> \cdots > x_{P(\N)}\}$.
For the sake of clarity, we shall focus on the fundamental sector $\mathcal{R}$.

The ground state (\ref{gsansatz}) is shown to be an eigenstate of the parent TCSM Hamiltonian 
\beqa
\label{HTCS}
\hat{\mathcal{H}}&=&
\sum_{i=1}^\N\!\Big[-\frac{\hbar^2}{2m}\frac{\partial^2}{\partial x_i^2}+\frac{1}{2}m\om^2x_i^2\Big]\\
& & +\sum_{\substack{i<j \\
|i-j|\leq r}}\frac{\hbar^2\lambda(\lambda-1)}{m|x_i-x_{j}|^2} + \sum_{\substack{i<j<k \\
|i-j|\leq r\\
|j-k|\leq r}}
\frac{\hbar^2\lambda^2}{m}\frac{{\bf r}_{ji}\cdot{\bf r}_{jk}}{{\bf r}_{ji}^2{\bf r}_{jk}^2},\nonumber
\eeqa
where ${\bf r}_{ij}=(x_i-x_j){\bf e}_x$ and ${\bf e}_x$ is a unit vector along the $x$ axis.  As a result, Hamiltonian (\ref{HTCS}) describes $\N$ particles harmonically confined in one dimension and interacting through a pairwise two-body potential as well as a three-body term. 
We note that the truncation of the interaction is not mediated by the relative distance between particles, but rather by the number of neighbors. As such, it represents a screening effect. The maximum range of the two-body interactions is $r$ while that of the three-body interactions is $2r$, as $r<|i-k| \leq 2r$. 
The two-body interactions are attractive for $\lambda\in[0,1)$ and repulsive for  $\lambda\geq 1$. By contrast, the three-body interactions are always attractive over $\mathcal{R}$.  Indeed, the three-body term can be conveniently rewritten as 
\beqa
 \sum_{\substack{i<j<k \\
|i-j|\leq r\\
|j-k|\leq r}}
\frac{{\bf r}_{ji}\cdot{\bf r}_{jk}}{{\bf r}_{ji}^2{\bf r}_{jk}^2}
 =-\sum_{\substack{i<j<k \\
r<|i-k|\leq 2r}}
\frac{1}{|x_i-x_j||x_j-x_k|}.
\eeqa
For $r=\N-1$ this term vanishes identically and one recovers the ground state of the Calogero-Sutherland model for indistinguishable bosons restricted to a sector \cite{Calogero71,Sutherland71}.
In particular, for $r=\N-1$ and $\lambda=1$ one recovers the Tonks-Girardeau gas in a harmonic trap, describing one-dimensional bosons with hard-core interactions \cite{Girardeau60,GWT01}.  To have $\N$ distinguishable particles in the limit $r=\N-1$, the inverse square interaction would need to be $\lambda^2-\lambda M_{ij}$, where $M_{ij}$ permutes particle coordinates \cite{ujino1998new}.  As a result, we shall refer to the case with $\lambda=1$  and $r<\N-1$ as a truncated Tonks Girardeau (TTG) gas.  For $\lambda=0$ the system describes an ideal Bose gas. 
In this work we focus on the case where $\lambda>0$, where full permutation symmetry is broken for $r<\N-1$.  For $r=1$, the Hamiltonian (\ref{HTCS}) reduces to the model with  nearest and  next-nearest neighbors discussed in  \cite{JK99,Ezung05}.
As a result,  the system described by the TCSM Hamiltonian (\ref{HTCS}) interpolates between the CSM and the Jain-Khare model.

The ground-state energy of the system is given by
\beqa
E_{\N,\lambda,r}^0=\frac{\hbar\om}{2}[\N+\lambda r(2\N-r-1)].
\eeqa
This expression suggests a mean-field theory where the zero-point energy of $\N$ non-interacting particles is shifted by $\lambda\hbar\om$ times the number of interacting pairs of particles, $r(2\N-r-1)/2$.
Hence, the zero-energy contribution  reproduces the well-known result for the full Calogero-Sutherland model when the range of the interactions is set to
$r=\N-1$.
In this limit, the scaling is quadratic in the particle number $\N$. Else, for $r<\N-1$, $E_{\N,\lambda,r}^0 \propto \N$ and 
depends quadratically on the range of the interaction $r$.

\section{Ground State Correlations}\label{SecCorr}

\begin{figure*}[t]
\centering
\subfigure{
\includegraphics[scale=.8]{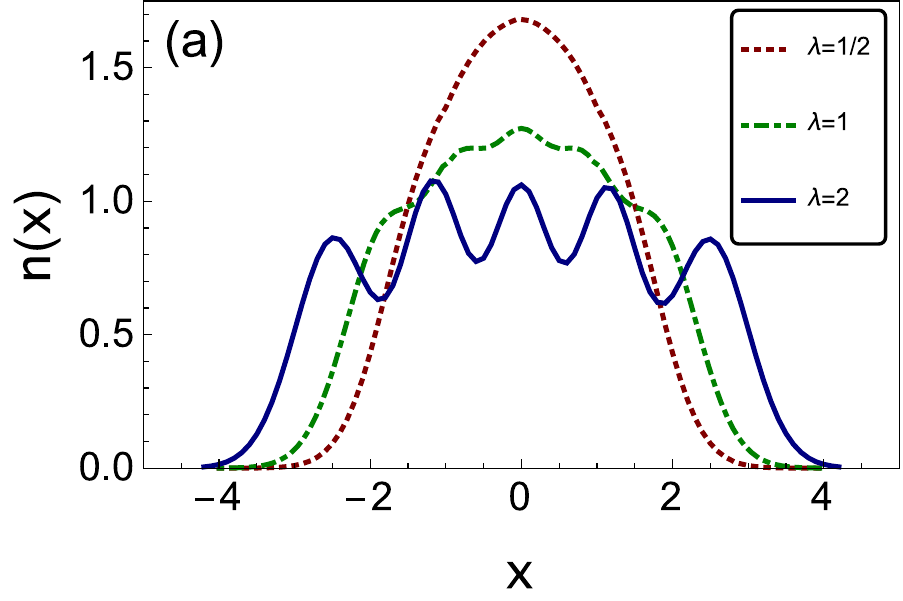}
\label{fig:N5r2density}
}
\subfigure{
\includegraphics[scale=.8]{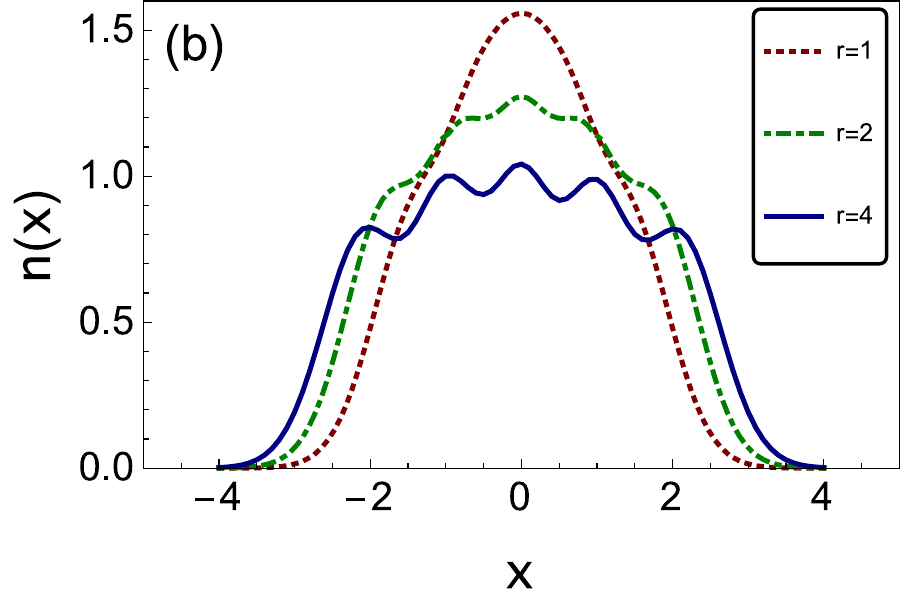}
\label{fig:N5ralldensity}
}
\caption{{\bf Local quantum correlations of the TCSM.} The density profile is plotted as a function of distance with respect to the center of the trap for $\N=5$. (a) Increasing the interaction strength while keeping the range fixed ($r=2$) leads to spatial antibunching reflected in the fringes of the density profile.  (b) For a given strength of the interactions ($ \lambda=1$), the TCSM interpolates between the Jain-Khare model ($r=1$) and the rational Calogero-Sutherland model ($r=\N-1$) as the range is increased, enhancing spatial antibunching. The case $ \lambda=1$ shown here can be thought of as a truncated Tonks-Girardeau (TTG) gas.}
\end{figure*}

\begin{figure*}[t]
\centering
\subfigure{
\includegraphics[scale=.42]{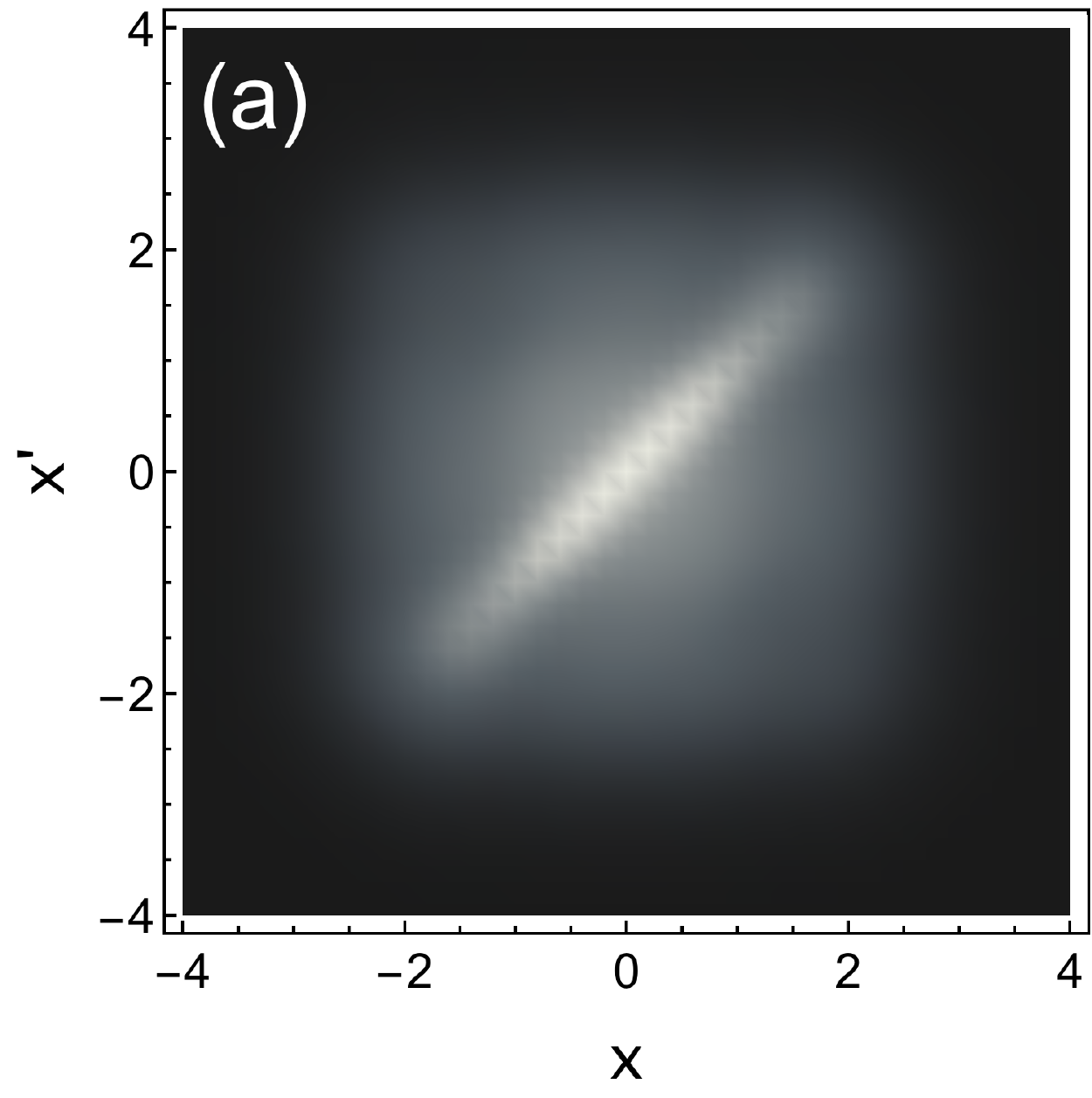}
\label{fig:N5r1densitymatrix}
}
\subfigure{
\includegraphics[scale=.42]{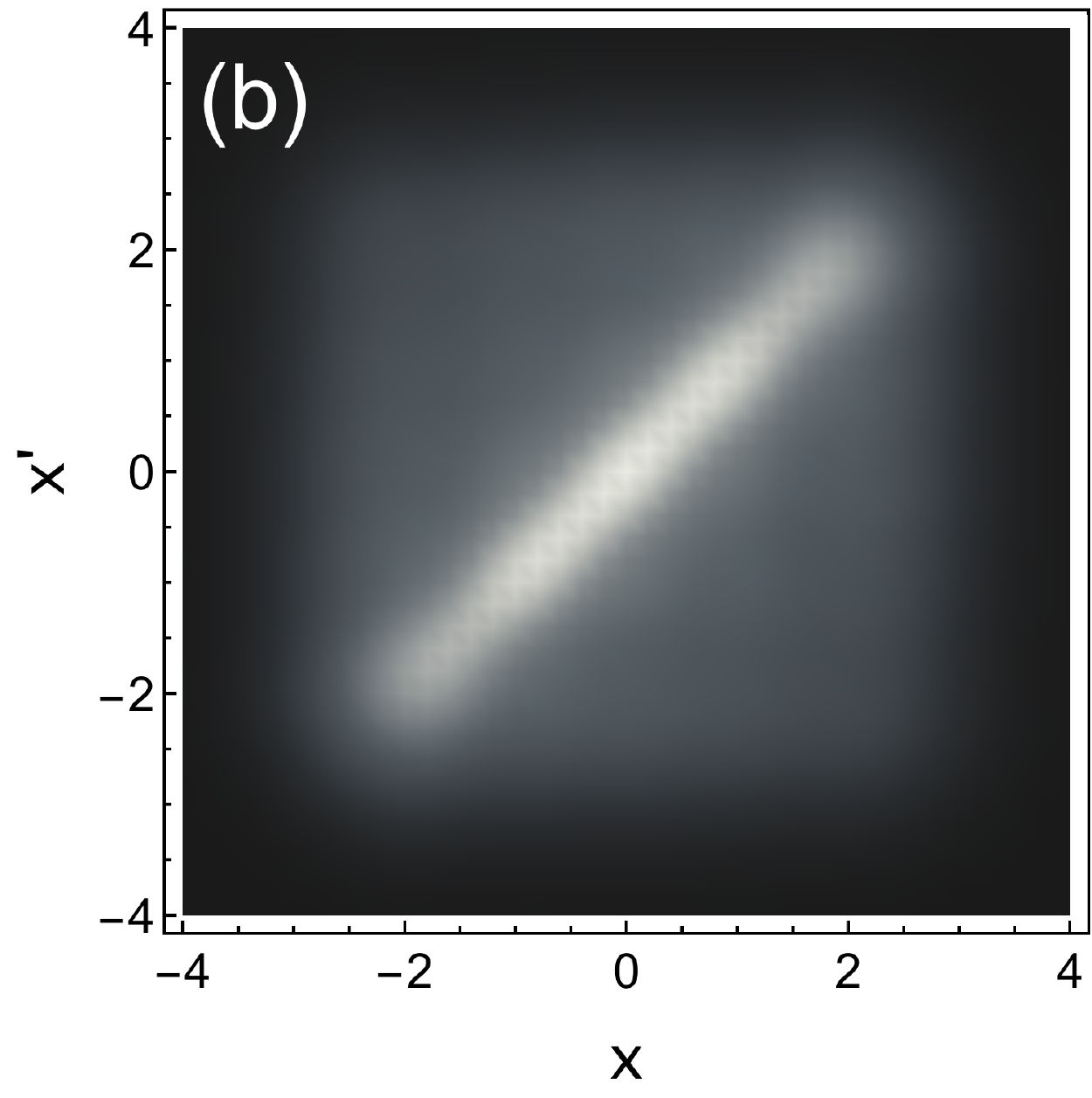}
\label{fig:N5r2densitymatrix}
}
\subfigure{
\includegraphics[scale=.42]{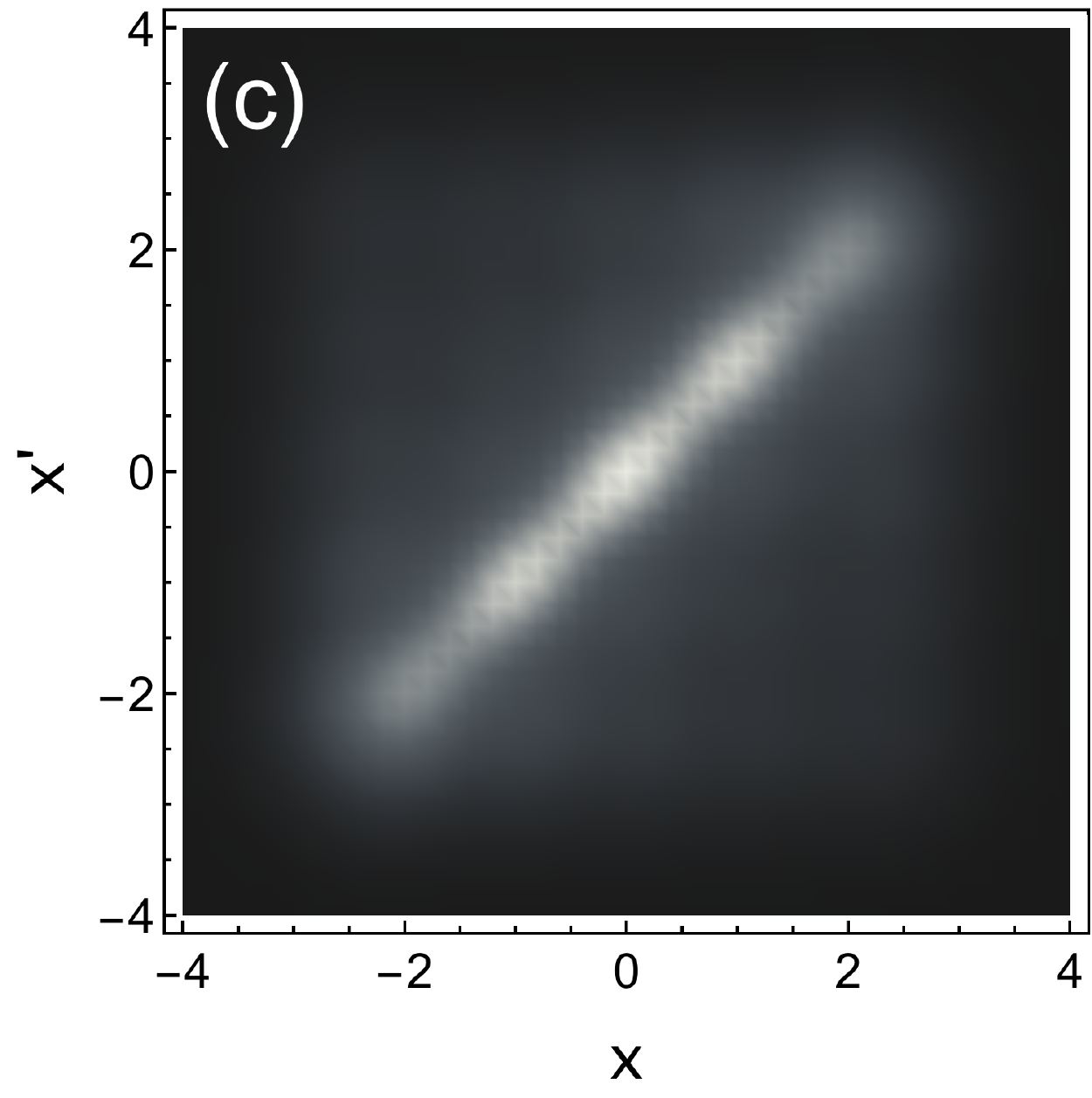}
\label{fig:N5r3densitymatrix}
}
\caption{{\bf One-body reduced density matrix of the symmetrized TCSM.} Plot of $\rho(x,x')$ as a function of space for $\N=5$ particles and interaction strength $\lambda=1$, i.e., the TTG gas. 
 The case $r=1$ in (a) is contrasted with $r=2$ in (b) and $r=3$  in (b), showing that  off-diagonal long-range order is suppressed as the interaction range $r$ increases.  The color coding runs from dark to light with increasing value of $\rho(x,x')$.}
\label{fig:densitymatrix}
\end{figure*}

\begin{figure*}[t]
\begin{center}
\includegraphics[height= 2 in]{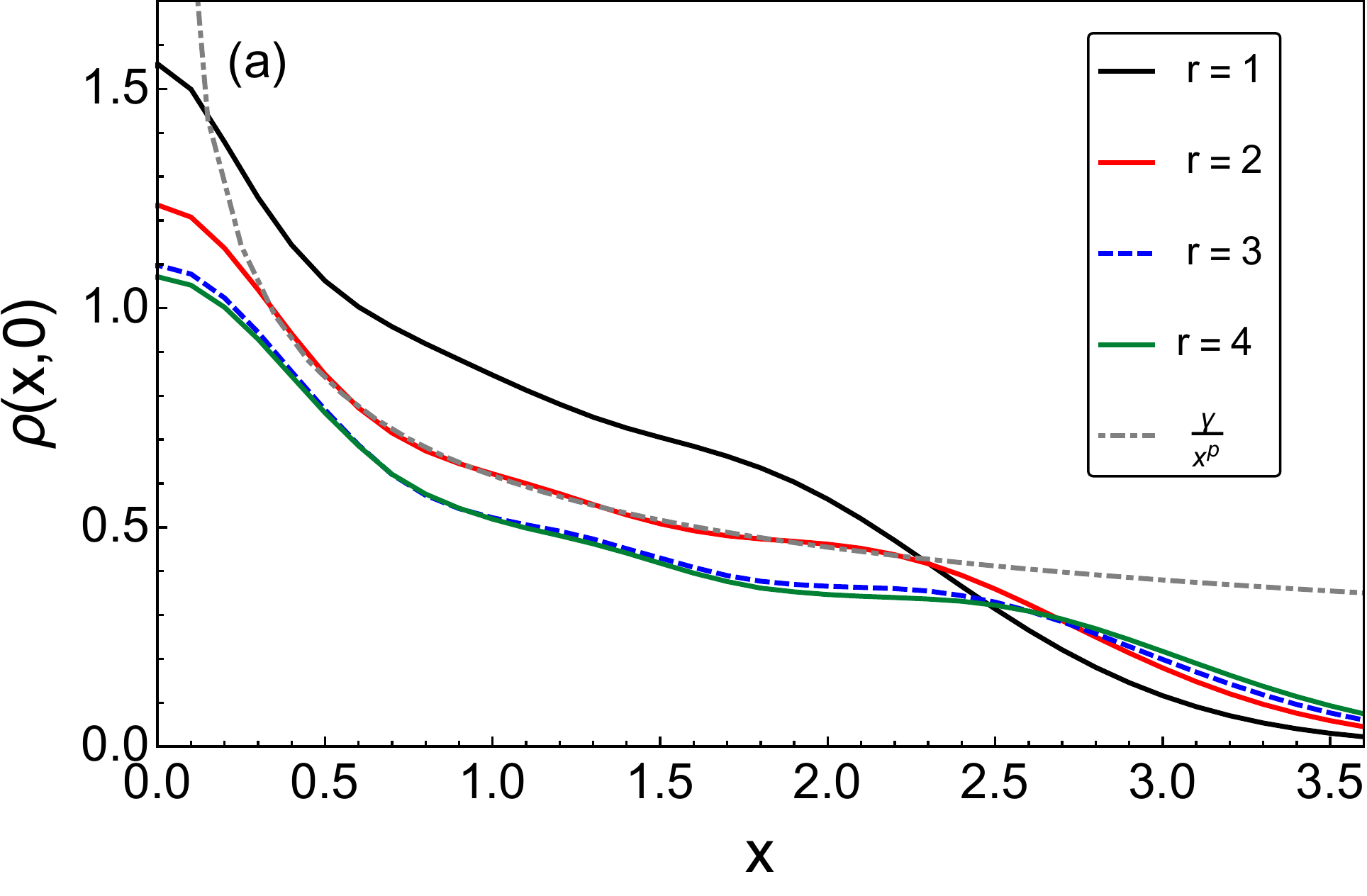}  
\hspace{0.5cm} 
\includegraphics[height= 2 in]{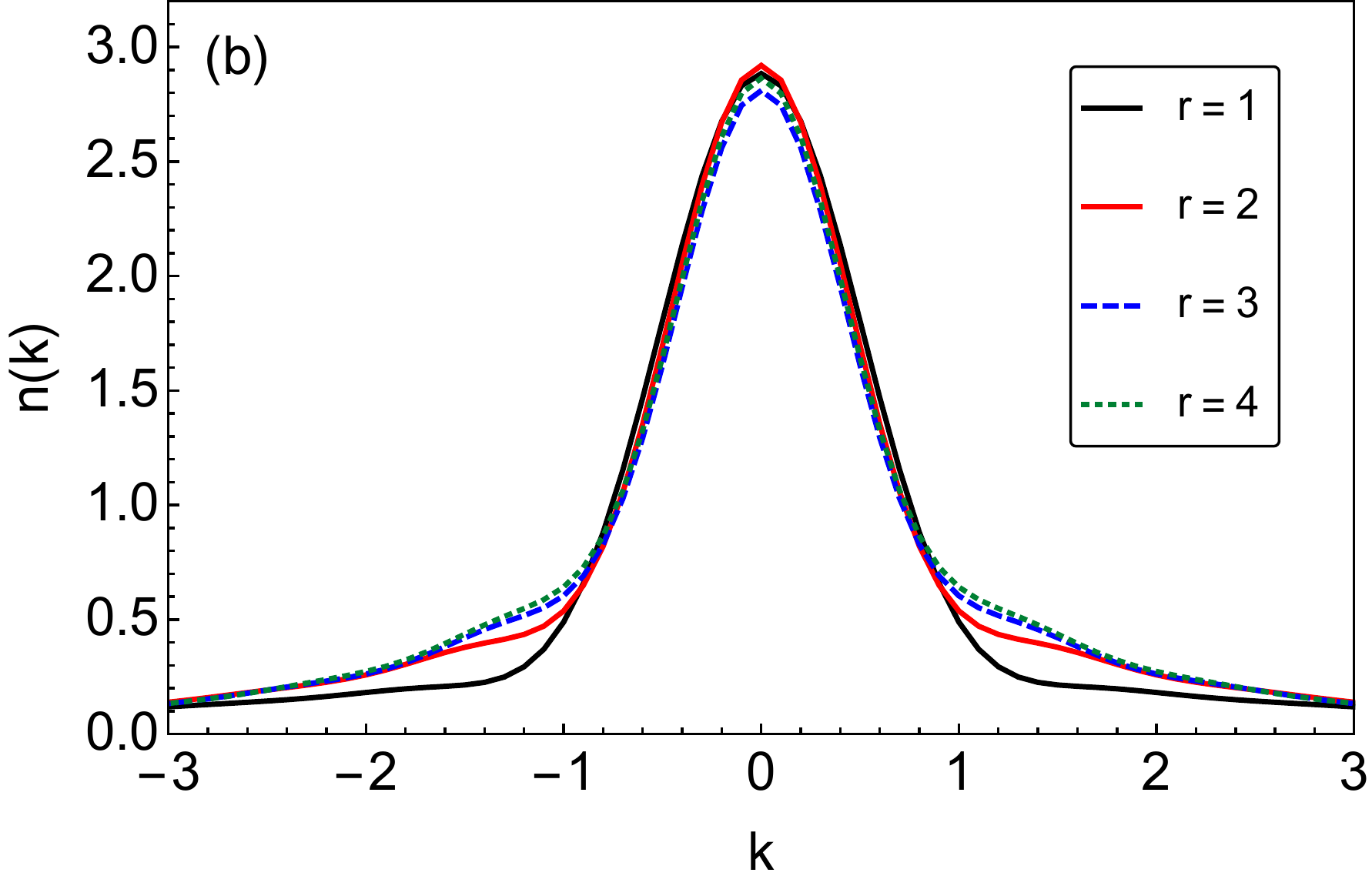}
\caption{ \textbf{Nonlocal quantum correlations of the TCSM ground state.} (a) Decay of the one-body reduced density matrix $\rho(x,0)$ as a function of the distance away from the center of the trap.  $\N=5$, $\lambda=1$ and $r=\{1,2,3,4\}$, from top to bottom.  Away from the origin, a power-law fit $\rho(x,0)=\frac{\gamma}{|x|^{p}}$, yields an exponent $p$ that increases with the interaction range $r$.  This is consistent with the loss of off-diagonal long-range order with increasing $r$. (b) Momentum distribution $n(k)$ for the symmetrized TTG. While the momentum distribution is sharply peaked at $k=0$, its tails are broadened as the range of the interaction increases until recovering the full-range model, the conventional TG gas with $r=\N-1$.}
\label{fig:OBRDMcentral}
\end{center}
\end{figure*}

The knowledge of the exact ground state of the Hamiltonian (\ref{HTCS}) allows us to investigate the role of the truncation of the inverse-square interactions.  
To this end, we next focus on the characterization of  one-body correlations. We distinguish local correlations that depend only on the diagonal elements of the density matrix in the position representation from non-local correlations that depend explicitly on off-diagonal coherences.
A prominent example of a one-body local correlation is the density profile which by its definition it is shared by the unsymmetrized and symmetrized TCSM.
And example of a one-body nonlocal correlation function is the momentum distribution, the Fourier transform of the one-body reduced density matrix (OBRDM) that explicitly depends on the coherences of the later. We therefore anticipate that nonlocal correlations will differ for the 
TCSM on a sector and its symmetrized version describing a quantum fluid of indistinguishable particles.

The fully symmetrized ground-state wavefunction reads,
\beqa
\label{wfgss}
\Psi^S_0(\mathbf{x})=\phi(\mathbf{x})\varphi^S(\mathbf{x}),
\eeqa
\noindent
where $\varphi^S(\mathbf{x})$ is given in Eq.  (\ref{symmetric}).  In particular, we consider the density profile of the the ground state, defined as
\beqa \label{density}  
n(x)&=&\N\int_{\mathbb{R}^{\N-1}} d^{\N-1}{\bf x}\, |\Psi^S_{0}(\mathbf{x})|^2, \nonumber \\
&=&\sum_{P_1\in\ \sigma^P}\int_{\mathbb{R}^{\N-1}} d^{\N-1}{\bf x}\,  \Big|\Psi_0\Big(P_1(\mathbf{x})\Big)\Big|^2 \Theta_{P_1}(\mathbf{x}),  
\eeqa
where $\mathbf{x}=(x,x_2\cdots x_\N)$,  $\sigma^P=\{(1\cdots p),  p=1,\cdots \N  \}$ is the set of $p$-cycles that effectively shifts $x$ to the right of each element in $\mathbf{x}$, and the identity $\Theta_P(\mathbf{x})\nonumber  \Theta_{P'}(\mathbf{x})=\Theta_P(\mathbf{x})\delta_{P,P'}$ was used. 
We emphasized that the density profile of the symmetrized TCSM  is shared by the the unsymmetrized model whose ground state is simply described by Eq. (\ref{gsansatz}), instead of (\ref{wfgss}).
For a fixed particle number $\N$ and range $r$, the density profile develops signatures of spatial antibunching as the interaction strength $\lambda$ is increased. This is confirmed by the numerical integration using the Monte Carlo method in Figure \ref{fig:N5r2density} for $\N=4,\ r=2$ for $\lambda=\{0,\frac{1}{2},1,2\}$. As $\lambda$ is increased, the density profile varies from a bell-shape function to a broader distribution that shows fringes and a number of peaks that equals the particle number $\N$.

When $\lambda=1$ and $r=\N-1$,  even in the absence of symmetrization,  the density profile  reduces to that of a  Tonks-Girardeau gas that describes one-dimensional bosons with hard-core interactions. In the ground state, its explicit form is $n(x)=\sum_{n=0}^{\N-1}|\phi_n(x)|^2$,  where $\phi_n(x)$ are the single-particle eigenstates of the harmonic oscillator. This expression is  identical to that of  the density profile of polarized and spinless fermions \cite{Girardeau60,GWT01}.   In the limit $r=0$ one recovers the ideal Bose gas, with $n(x)=\N |\phi_0(x)|^2$.
The TCSM with $\lambda=1$ is equivalent to  a TTG gas and interpolates between these two limits for $0<r<\N-1$.  The visibility of the fringes in $n(x)$ diminishes then as  the range of the interactions is decreased, see Fig. \ref{fig:N5ralldensity} for $\N=5$ and  $r=\{1,2,4\}$.

We next turn our attention to the characterization of off-diagonal correlations.
In particular, we consider the one-body reduced density matrix, that for the symmetrized model is defined as
\begin{eqnarray}
 & & \rho(x,x') = \N\int_{\mathbb{R}^{\N-1}} d^{\N-1}{\bf x}\, \Psi^S_{0}(\mathbf{x})\Psi^{*S}_{0}(\mathbf{x}'), \nonumber \\
 & &= \sum_{P_1,P'_1\in\ \sigma^P}\int_{\mathbb{R}^{\N-1}} d^{\N-1}{\bf x}\,  \Big[ \Psi_0\Big(P_1(\mathbf{x})\Big) \Psi^*_0\Big(P'_1\circ P_1(\mathbf{x}')\Big)\nonumber \\
& &\  \ \ \ \ \ \ \ \ \ \ \ \times\, \Theta_{P_1}(\mathbf{x}) \Theta_{P'_1 \circ P_1}(\mathbf{x}')\Big],
 \label{simplematrix}
\end{eqnarray}

\noindent
where $\mathbf{x}'=(x',x_2,\cdots,x_\N)$  and the expression in the second line is suitable for Monte Carlo integration. Here, $\sigma^P$ is as previously defined.  Figure~\ref{fig:densitymatrix} shows $\rho(x,x')$ for (a) $\N=5,\lambda=1,\ r=1$, (b) $\N=5,\lambda=1,\ r=2$, and (c) $\N=5,\lambda=1,\ r=3$.  The maximum amplitude lies along the diagonal, and matches the density profile, e.g,  $\rho(x,x)=n(x)$.  The amplitude decreases when going away from the diagonal.  This characterizes a loss of off-diagonal long range order,  that decays  much faster for $r=3$ than $r=1$.  This is consistent with the fact that interactions are suppressed as  $r$ is decreased, approaching the ideal Bose gas for $r=0$. 
The $r=3$ system is closer to the Tonks-Girardeau gas describing impenetrable bosons in a harmonic trap, where the off-diagonal long range order has been shown to vanish in the thermodynamic limit \cite{GWT01}.  

Figure~\ref{fig:OBRDMcentral} shows the OBRDM for $\N=5,\lambda=1$ and $r=\{1,2,3,4\}$ with varying distance from the trap center $x$ and $x'=0$, which exhibits algebraic decay away from the center $\rho(x,0)=\frac{\gamma}{x^{p}}$.  Table~\ref{tab:powerlawden} shows the numerically calculated parameters for each case,  As expected, the exponent $p$ increases with increasing $r$.  The exponent for $r=3$ is $p=0.49\pm0.01$, which is the nearly the same as the finite size harmonically trapped Tonks-Girardeau gas $\rho(x,0)\sim \frac{1}{|x|^{1/2}}$~\cite{gangardt2004universal} (e.g. the $r=4$ case).  This is to be expected as the $r=3$ case has only one less interacting pair than the full CSM.  

\begin{table}
\caption{Parameters for numerical fit of OBRDM, $\rho(x,0)=\frac{\gamma}{|x|^{p}}$}
\begin{center}
    \begin{tabular}{c  c  c}
    \hline
    \hline
   \ \ \ $r$\ \ \  &\ \ \ \ \ $\gamma$\ \ \ \ \  &\ \ \ $p$ \ \ \ \\
   \hline
   1 &\ \ $0.830\pm 0.004\ \  $ & $0.354\pm 0.008$ \\
   2 & $0.617 \pm 0.002 $ & $0.442\pm 0.006$ \\
   3 & $0.527\pm 0.004$ & $0.49\pm 0.01$ \\
   4 & $ 0.521 \pm 0.003 $&  $0.50\pm 0.01$  
 \end{tabular}
\label{tab:powerlawden}
 \end{center}
 \end{table}

The momentum distribution is defined,
\begin{eqnarray}
n(k)=\frac{1}{2\pi}\int^{\infty}_{-\infty}dx \int^{\infty}_{-\infty} dx' \rho(x,x')e^{-i k (x-x')}
\end{eqnarray}
\noindent
and is normalized to $\int^{\infty}_{-\infty} n(k)dk=\N$.  Figure~\ref{fig:OBRDMcentral}(b) shows the numerically calculated momentum distribution  $n(k)$ versus $k$ in the symmetrized system for $\N=5$, $\lambda=1$, and $r=\{1,2,3,4\}$.   The momentum distributions feature a central peak at $n(k=0)$ similar to the ideal Bose gas case (e.g. $r=0$).  The tails become increasingly broad as the interaction range $r$ is increased.

\section{Collective Excitations Above the Ground State}
A set of collective excitations above the ground state  can be found using the ansatz $\Psi=\Psi_0\Phi$. As a result of the factorized form of $\Psi$, the time-independent Schr\"odinger equation, $\hat{\mathcal{H}}\Psi=E\Psi$, reduces to the eigenvalue equation
\beqa
(2 \tilde{\om} \hat{K}-\hat{D}_+)\Phi=\varepsilon\Phi,
\eeqa
where $\tilde{\om}=\frac{m\omega}{\hbar}$, $\varepsilon=\frac{2m}{\hbar^2}(E-E^0_{\N,r,\lambda})$, and
\beqa
\hat{K}&=&\sum_{i=1}^\N x_i\frac{\partial}{\partial x_i},\\
\hat{D}_+&=&\sum_{i=1}^\N \frac{\partial^2}{\partial x_i^2}+\sum_{\substack{i<j \\
|i-j|\leq r}}\frac{2\lambda}{x_i-x_j}\Big(\frac{\partial}{\partial x_i}-\frac{\partial}{\partial x_j}\Big),
\eeqa
\noindent
that satisfy
\beqa
[\hat{D}_{+},\hat{K}]=2\hat{D}_{+}.
\eeqa  
\noindent
Defining $\hat{K}'=-\frac{1}{2}\Big(\hat{K}+\frac{E_{\N,\lambda,r}^0}{\hbar \omega}\Big)$, $\hat{D}'_{+}=\frac{1}{2}\hat{D}_+$, and $\hat{D}'_{-}=\frac{1}{2}\sum_i x^2_i$, these operators $\{\hat{K}',\hat{D}'_{\pm}\}$ are the generators of  the SU$(1,1)$ algebra, 
\beqa
[\hat{D}'_{+},\hat{D}'_{-}] =-2\hat{K}',\ \ \  [\hat{K}' ,\hat{D}'_{\pm}] = \pm \hat{D}'_{\pm}.
 \eeqa
Under the action of the similarity transformation $\hat{A}=\Psi_0\exp(-\hat{D}_+/4\tilde{\omega})$,  collective excitations of the  
TCSM Hamiltonian (\ref{HTCS}) are related to that of the  Euler operator $\hat{K}$,
\begin{eqnarray}
\label{transformedham}
\hat{A}^{-1}(\hat{\mathcal{H}}-E^{0}_{\N,r,\lambda})\hat{A} =\hbar \omega \hat{K}. 
\end{eqnarray}
\noindent
An eigenstate of $\hat{K}$ is a homogeneous symmetric monomial $S_n$  with eigenvalue equal to the degree  $n$, $\hat{K}S_n= n S_n$.

The TCSM Hamiltonian can be further simplified by the additional transformation,
\begin{eqnarray}
\label{decouple}
\phi(\mathbf{x})\hat{A}_{0}\hat{A}^{-1}(\hat{\mathcal{H}}-E^0_{\N,r,\lambda})\hat{A}\hat{A}^{-1}_{0}\phi^{-1}(\mathbf{x}) =\nonumber \\
-\frac{\hbar^2}{2m}\sum_{i=1}^\N \frac{\partial^2}{\partial x_i^2} +\frac{m\omega^2}{2}\sum_{i=1}^{\N}x^2_i - \frac{\hbar \omega}{2} \N,
\end{eqnarray}
where $\hat{A}_{0}=\exp\Big[-\hat{D}_{+,\lambda=0}/4\tilde{\omega}\Big]$ and $\phi(\bf{x})$ is the  Gaussian function given by Eq.  (\ref{gaussian}).  This particular family of states can thus be mapped to that of $\N$ decoupled harmonic oscillators. For $r=\N-1$, it is well-known that the excitation spectrum and level degeneracy of the interacting system is equivalent to that of noninteracting bosons in a harmonic trap \cite{Kawakami93,GP99}.  For $r<\N-1$, the singular terms in the similarity transformation can lead to nonpolynomial solutions, ~\cite{BMK01,Ezung05} suggesting the quasi-exactly solvable nature of this model. 

One approach to find collective excitations above the ground state uses the operator $\hat{A}$ in an analogous way as discussed in Ref.~\citenum{Sogo96} for the full-range CSM ( $r=\N-1$) and in Ref.~\citenum{Ezung05} for the Jain-Khare model ($r=1$).   The excitations with energy eigenvalue $\hbar \omega n$ are given by $\Psi_n=\Psi_0\Phi_n$ with
\begin{eqnarray}
\label{similarityoperator}
\Phi_n=\exp(-\hat{D}_{+}/4\tilde{\omega})S_n.
\end{eqnarray}
\noindent
However, for $r<\N-1$ some of the excited states found using this method may not be normalizable (e.g. will not end as a polynomial) for a given energy level $n$ and homogeneous symmetric monomial $S_n$.  A similar result was found for the Jain-Khare model~\cite{Ezung05}.

Collective excitations can alternatively be found using Calogero's approach~\cite{Calogero71}, which separates (\ref{HTCS}) into radial and angular parts and finds the solutions for each using the form,
\begin{eqnarray} 
\label{full}
\Psi(\mathbf{x})=\varphi(\mathbf{x}) \Phi_{n,k}(\rho^2)P_{k}(\mathbf{x}),
\end{eqnarray}
\noindent
where $\rho^2=\sum_{i=1}^\N x^2_i$ is the radial degree of freedom and $\varphi(\mathbf{x})$ is given in Eq.  (\ref{zeqn}).  Here $P_k(\mathbf{x})$ is a symmetric homogeneous polynomial of degree $k$ that satisfies a generalized Laplace equation, 
\beqa
\label{LaplaceEq}
\hat{D}_+P_k(\mathbf{x})=0.
\eeqa
  The radial solution reads
\begin{eqnarray}
\Phi_{n,k}(\rho^2)=\exp(-\tilde{\om}\rho^2/2)L^{\nu}_n({\tilde{\omega} \rho^2}),
\end{eqnarray}
\noindent
where $\nu=\frac{E^{0}_{\N,\lambda,r}}{\hbar\om}+k-1$ and $L^{\nu}_n(\tilde{\omega} \rho^2)$ is a generalized Laguerre polynomial
\beqa
L^{\nu}_n(\tilde{\omega} \rho^2)=\sum^n_{m=0}\binom{\nu +n}{n-m}(-1)^m\frac{(\tilde{\om}\rho^2)^m}{m!},
\eeqa
with  $n\in\mathbb{N}_0$.  The generalized Laguerre polynomial, as well as the coefficient $\nu$, can also be found from (\ref{similarityoperator}), by taking $S_{(2n+k)} \propto \rho^{2n}P_k(\mathbf{x})$ and using the following relation~\cite{Ezung05}, 
\begin{eqnarray}
\frac{\hat{D}_+}{4\tilde{\omega}}\Big[\rho^{2n}P_{k}(\mathbf{x})\Big]= \frac{n}{\tilde{\omega}}\Big[\frac{E^0_{\N,\lambda,r}}{\hbar\om}+k-1+n\Big]\rho^{2(n\text{-}1)}P_{k}(\mathbf{x}).\nonumber\\
\end{eqnarray} 
The corresponding excitation energy is linear in the quantum numbers $n$ and $k$, 
\beqa
(E-E^{0}_{\N,\lambda,r})=\hbar \om(2n+k) =\hbar \om s,
\eeqa
for $s\in\mathbb{N}_0$, as anticipated from (\ref{decouple}).  Such result extends to  variations of the CSM preserving SU$(1,1)$ in one spatial dimension~\cite{Calogero71,Gambardella75,ASK01,BMK01,MMS03}.  The advantage of this approach is that the level degeneracy within this family of solutions is more transparent.

\begin{table*}[t]
\begin{center}
\begin{tabular}{c  l l  }
\hline
\hline
$k$\ \  &\ \ $ P_k(\mathbf{x})$  &  Coefficients $\{ c_\alpha\}$ \\
\hline
$k=1$\ \  &\ \  $P_{1}(\mathbf{x})=c_1 m_1(\mathbf{x})$ & N/A \\
\\
$k=2$ \ \ &\ \  $P_2(\mathbf{x})=c_{2}m_{2}(\mathbf{x})+c_{11}m_{11}(\mathbf{x})$ & $\lambda=\frac{ 2  \mathrm{N} c_2}{(c_{11} - 2 c_2)r(2 \mathrm{N}-r-1)}$ \\
\\
$k=3$\ \ & \ \ $P_3(\mathbf{x})=c_{3}m_{3}(\mathbf{x})+c_{21}m_{21}(\mathbf{x})+c_{111}m_{111}(\mathbf{x})$ &  $\lambda =\frac{  2 \Big[3 c_3 + c_{21} (\mathrm{N} -1) \Big]}{(c_{21} - 3 c_3) r (2\mathrm{N}-r-1)}$ \\
& & $c_{111}= 3 (c_{21} - c_3)$ \\
\\
$k=4$\ \ &\ \  $P_4(\mathbf{x})=c_{4}m_{4}(\mathbf{x})+c_{31}m_{31}(\mathbf{x})+c_{22}m_{22}(\mathbf{x})$ & $\lambda=\frac{2 \Big[ 6 c_4 + c_{22} (\mathrm{N}-1)\Big]}{(c_{31} - 4 c_4) r( 2 \mathrm{N} - r-1)}$ \\
&\ \ \ \ \ \ \ \ \ \  $+c_{211}m_{211}(\mathbf{x})+c_{1111}m_{1111}(\mathbf{x})$ &  $c_{1111} = 6 (c_{22} - 2 c_4)$ \\
& & $c_{211} = c_{31} + 2 c_{22} - 4 c_4$ \\
& & $\Big[(\mathrm{N}+4)c_{31}+(\mathrm{N}-2)(2c_{22}-4c_4) $ \\
& & \ \ \  $+ \lambda r(2\mathrm{N}-r-1) (2c_4+c_{31}-c_{22})\Big]=0$ \\ 
\\
$k=5$\ \ &\ \ $P_5(\mathbf{x})=c_{5}m_{5}(\mathbf{x})+c_{41}m_{41}(\mathbf{x})+c_{32}m_{32}(\mathbf{x})$ & $\lambda=\frac{2 \Big(10 c_5 + ( \mathrm{N}-1)c_{32}\Big)}{(c_{41} - 5 c_5) r (2\mathrm{N}-r-1)}$ \\
& \ \ \ \ \ \ \ \ \ \ $+c_{311}m_{311}(\mathbf{x})+c_{221}m_{221}(\mathbf{x})+c_{2111}m_{2111}(\mathbf{x}) $\ \ \ \ \ \ \  & $c_{311} = 2 c_{32} + c_{41} - 5 c_5$ \\
& \ \ \ \ \ \ \ \ \ \ $+c_{11111}m_{11111}(\mathbf{x})$ & $c_{221} = 5 c_{32} - 3 c_{41} - 5 c_5$ \\
 & & $c_{2111} = 3 (4 c_{32} - 3 c_{41} - 5 c_5)$ \\
 & & $c_{11111} = 30 (c_{32} - c_{41} - c_5)$ \\
 & & $\Big[(5 \mathrm{N}  - 7)c_{32}-3 (\mathrm{N} - 4) c_{41}-5 (\mathrm{N} - 2) c_5 $  \\ 
 & &\ \ \  $+ \frac{\lambda r}{2} (2 \mathrm{N} - r - 1) (5 c_5+3c_{41}-2 c_{32})\Big]=0 $ \\
 \hline
\end{tabular}
\caption{Excited states of the TCSM are derived  via (\ref{full}) in terms of a symmetric homogeneous polynomials $P_k(\mathbf{x})$ of degree $k$, constructed as linear combination of symmetric monomial functions  (\ref{monomialfunction}) with  coefficients $c_\alpha$, that satisfy the listed constraints for $r<\mathrm{N}-1$ and $k\leq \N$.}
\label{Coefficients}
\end{center}
\end{table*}

\begin{table*}[t]
\begin{center}
\begin{tabular}{c l }
 \hline	
 \hline
$k$\ \  & \ \ Constraints for $r=\mathrm{N}-1$ \\
\hline
$k=3$\ \ &\ \ $6c_3+2c_{21}(\mathrm{N}-1)+2\lambda(\mathrm{N}-1)\Big[3c_3-c_{21}+(\mathrm{N}-2)(c_{21}-\frac{c_{111}}{2})\Big]=0$ \\
\\
$k=4$\ \ &\ \ $12 c_4 + 2(\mathrm{N}-1)c_{22}+2\lambda (\mathrm{N}-1)\Big[4 c_4-c_{31}+(\mathrm{N}-2)(c_{22}-\frac{c_{211}}{2})\Big] =0$ \\
&\ \ $12 c_{31}+2(\mathrm{N}-2)c_{211}+4\lambda \Big[ 2 c_4 -c_{22}+(3\mathrm{N}-5)c_{31}\Big] +2\lambda(\mathrm{N}-2)\Big[(\mathrm{N}-5)c_{211}-\frac{(\mathrm{N}-3)}{2}c_{1111}\Big]=0$ \\
\\
$k=5$\ \ &\ \ $5 c_5 + c_{32} (\mathrm{N} - 1) + \lambda \Big[(5 c_5 - c_{41}) (\mathrm{N} - 1) + (c_{32} - c_{311}/2) (\mathrm{N} - 1) (\mathrm{N} - 2)\Big]=0$ \\
 &\ \ $9 c_{311} + (\mathrm{N}-3) c_{2111} +  \lambda \Big[3(4 c_{41} - 2 c_{221} + 2 c_{311}) + (9 c_{311} -3 c_{2111}) ( \mathrm{N} - 3)+ (c_{2111} - \frac{c_{11111}}{2}) (\mathrm{N}-4) (\mathrm{N}-3)\Big] =0$ \\
 &\ \ $6 c_{41} + 3 c_{32} + c_{221} (\mathrm{N} - 2) +  \lambda \Big[(5 c_5 + 3 c_{41} - 2 c_{32}) + (\mathrm{N} - 2)(4 c_{41} - c_{311}+3 c_{32} +(\mathrm{N}-4)c_{221} - \frac{c_{2111}}{2}( \mathrm{N} - 3) \Big]=0$\\
 \hline
\end{tabular}
\caption{Excited states of the CSM in a sector.  Constraints on coefficients $c_\alpha$ in (\ref{monomialfunction}) for the case $r=\mathrm{N}-1$ and $k\leq \N$ with all-to-all pairwise interactions, when the  three-body term vanishes.}
\label{Coefficients2}
\end{center}
\end{table*}

Here we explicitly calculate the first few $P_k(\mathbf{x})$, though in principle this entire family of states can be found using this method.  We construct $P_k(\mathbf{x})$ as a linear combination of symmetric monomial functions \cite{macdonald1998symmetric},
\begin{eqnarray}
P_k(\mathbf{x})=\sum_{\alpha}c_{\alpha}m_{\alpha}(\mathbf{x}),
\label{monomialfunction}
\end{eqnarray}
\noindent
where $\alpha=(\alpha_1,\alpha_2,\cdots,\alpha_\N)$ describes a partition where the parts $\alpha_i$ are positive integers listed in decreasing order $\alpha_1\geq \alpha_2\cdots \geq \alpha_\N\geq 0$.   The monomial functions are given by the sum over all distinct permutations $S_\alpha$ of $\mathbf{x}^\alpha=x_1^{\alpha_1}\cdots x_\N^{\alpha_\N}$,  
\begin{eqnarray}
m_{\alpha}(\mathbf{x})=\sum_{\sigma\in S_\alpha}x^{\alpha_1}_{\sigma(1)}x^{\alpha_2}_{\sigma(2)}\cdots x^{\alpha_\N}_{\sigma(\N)}.
\end{eqnarray}
\noindent
The weight of a partition $|\alpha|=k=\sum^\N_{\substack{i=1}}\alpha_i$ corresponds to the order of $P_{k}(\mathbf{x})$.  Each $P_k(\mathbf{x})$ has a total of $M(k)$ different coefficients, where $M(k)$ is the multiplicity of distinct partitions with a given weight $|\alpha|=k$. The requirement for $P_k(\mathbf{x})$ to satisfy the Laplace equation (\ref{LaplaceEq}) will place constraints on these coefficients.  

When solving for the constraints, there are four different cases to consider that depend on $r, k,$ and $\N$.  For $k\leq \N$, the partition length $l(\alpha)\leq \N,\forall \alpha$ with $|\alpha|=k$, and general expressions for the constraints on $c_\alpha$ can be found for all $r$.  However, for $k>\N$ the partition length $l(\alpha)> \N$ for some $|\alpha|=k$.  In this case, there is no general form for the constraints as they will depend on $\N,r$ and $k$ and will not include $\{c_\alpha\}$ where the partition length $l(\alpha)> \N$.  We do not provide the constraints for $k>\N$, but they can be found using the same approach as for $k\leq \N$.  
  
In table~\ref{Coefficients}, we list the constraints on the coefficients $c_\alpha$ in (\ref{monomialfunction}) for $r<\N-1$ and $k\leq \N$.  For $k\geq 3$, the level degeneracy is different in the case of $r<\N-1$ and $r=\N-1$.  We first consider the former case.

\noindent

\begin{figure*}[t]
\begin{center}
\includegraphics[height= 1.44 in]{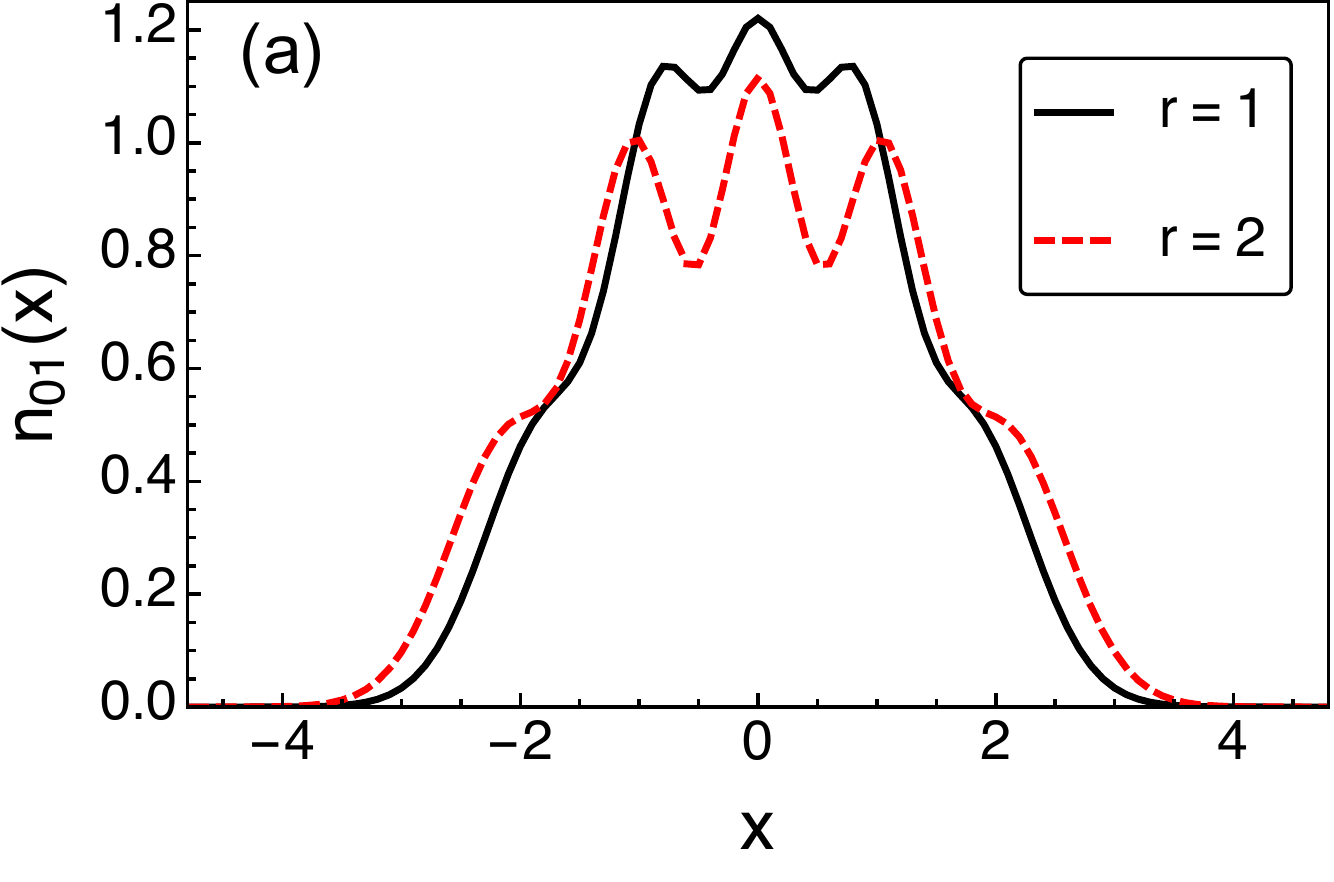} 
\hspace{0.2cm} 
\includegraphics[height= 1.44 in]{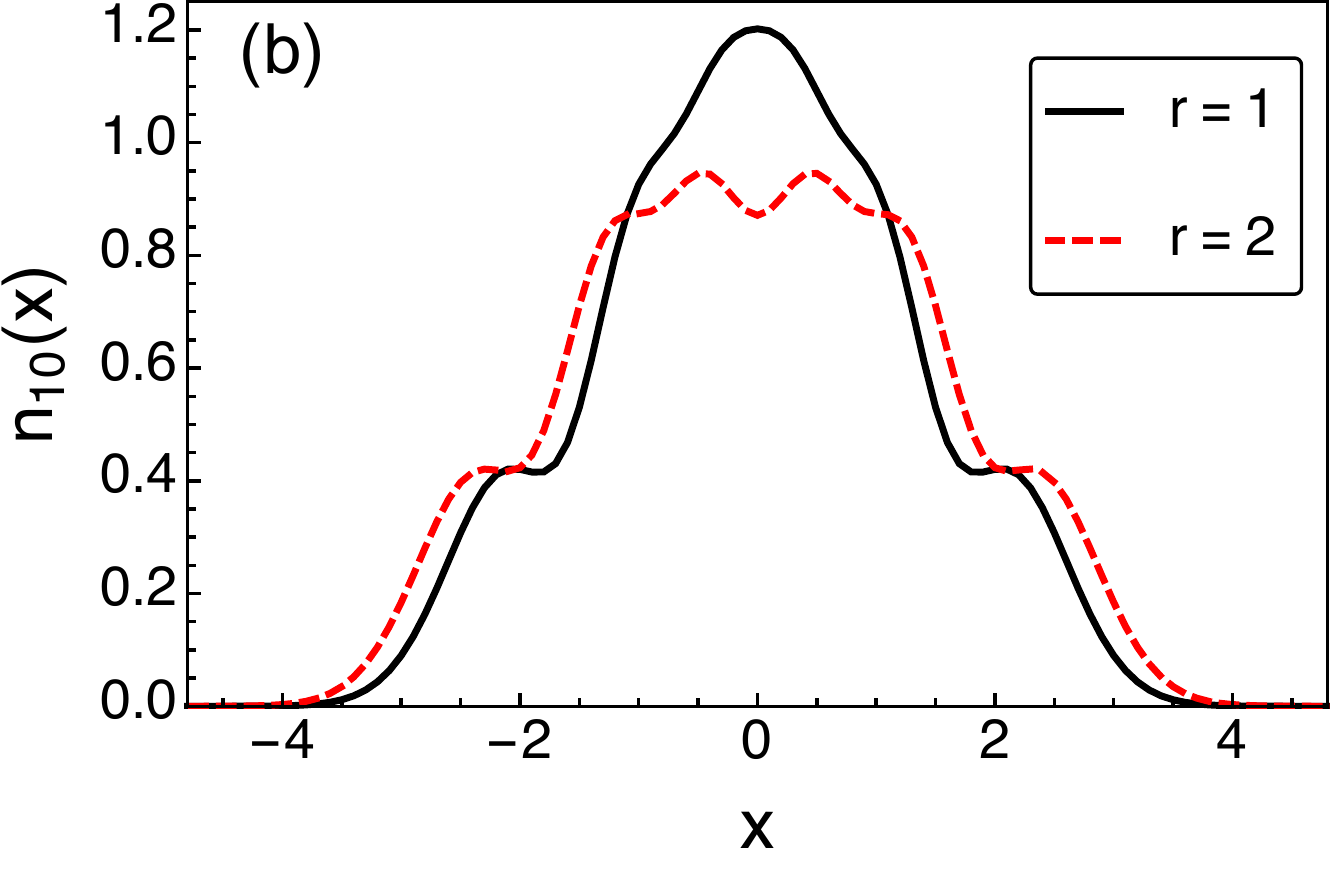} 
\hspace{0.2cm} 
\includegraphics[height=1.44 in]{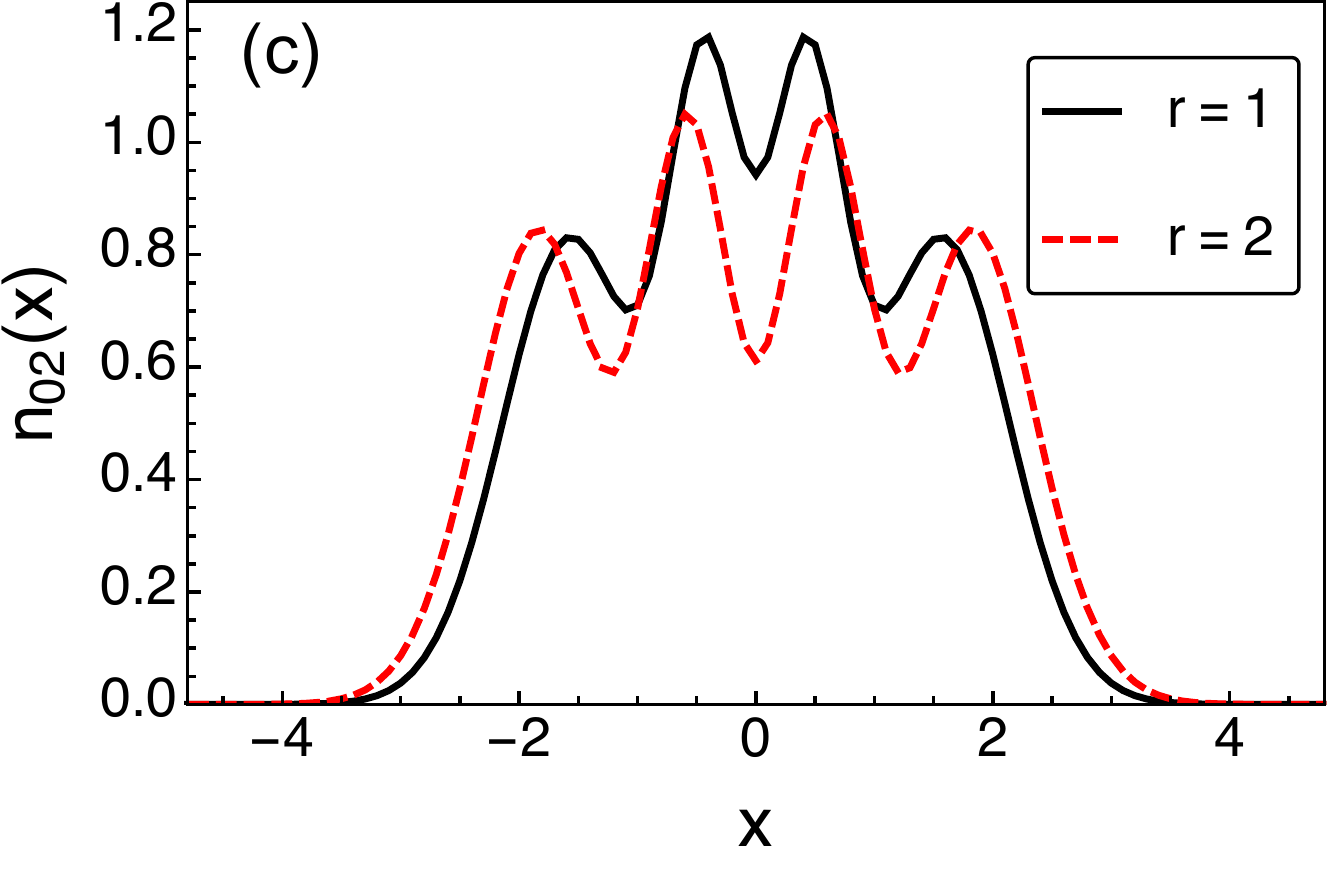}  
\caption{ \textbf{Density profile of excited states.}  These three plots show $n_{nk}(x)$ for $\N=4$ and $\lambda=1$ of the first three excited states: (a) $(n,k)=(0,1)$, (b) $(n,k)=(1,0)$, and (c) $(n,k)=(0,2)$.  In these three cases, increasing $r$ both increases the spatial extent of the distribution and the degree of spatial antibunching, leading to a higher visibility of the fringes in the density profile.  Such behavior is also observed in the ground state density $n(x)$ with increasing $r$, as shown in Fig.~\ref{fig:N5ralldensity}. }
\label{fig:}
\end{center}
\end{figure*}

For $r=1$, these expressions reproduce the constraints presented in~\cite{ASK01} for the Jain-Khare model.  For $r<\N-1$, there are $\mathcal{N}_c=M(k)-1$ constraints on the coefficients of each $P_k(\mathbf{x})$.  Consequently, $P_k(\mathbf{x})$ is a one-parameter family of solutions for each $k$ and $r<\N-1$.  The normalization requirement of each excited state yields only one distinct solution for $P_k(\mathbf{x})$ for a given $k$, and $(n,k)$ are the corresponding quantum number that describe each quantum state.  

The corresponding level degeneracy for each $s=2n+k$ reads
\begin{eqnarray}
\begin{split}
d(s)&=\frac{s}{2}+1,\ \  s=0,2\cdots 2p, \\
d(s)&=\frac{s+1}{2},\ \ s=1,3,\cdots 2p-1. 
\label{degeneracys}
\end{split}
\end{eqnarray}
\noindent
One finds the same degeneracy structure using an operator approach.  As we show in the {Appendix\ref{Appendix}}, performing a similarity transformation on the full Hamiltonian with $\varphi(\mathbf{x})$ (\ref{zeqn}) and separating the center of mass $X$ and relative degrees of freedom, the effective Hamiltonian describes two uncoupled oscillators,
\begin{eqnarray}
\varphi(\mathbf{x})^{-1} &\hat{\mathcal{H}}& \varphi(\mathbf{x})=\hat{\mathcal{H}}_{CM}+\hat{\mathcal{H}}_r, \\
\hat{\mathcal{H}}_{CM} &=& \frac{\hbar \omega}{2} \{a^{-}_X ,a^{+}_X \}= \hbar{\omega}\Big(\hat{N}_{X}+\frac{1}{2}\Big), \\
\hat{\mathcal{H}}_{rel} &=& \hbar \omega [a^{-}_{rel} ,a^{+}_{rel}] =\hbar{\omega}\Big(2\hat{N}_{rel}+\frac{E^{0,rel}_{\N,\lambda,r}}{\hbar \omega}\Big).
\end{eqnarray}
\noindent
Here $X=\frac{1}{\N}\sum^\N_{i=1}x_i$ is the center of mass and $\zeta_i=x_i-X$ are the relative degrees of freedom, and the creation and annihilation operators are
\begin{eqnarray}
a^{\pm}_X &=& \frac{1}{\sqrt{2}}\Big(\sqrt{\N\tilde{\omega}}X \mp \frac{1}{\sqrt{\N\tilde{\omega}}}\frac{\partial}{\partial_X} \Big), \\
a^{\pm}_{rel}&=&\frac{1}{2}\Big(\frac{ D'_{+}}{\tilde{\omega}}+\tilde{\omega} D'_{-} \Big)\pm K' ,
\end{eqnarray}  
\noindent
where the center of mass has now been factored out of $D'_{\pm}$ and $K'$.

As shown in the Appendix, these operators can be used to construct generators of the $\mathfrak{sl}(2,\mathbb{C})$ algebra for $s=2n_{r}+n_X\geq 2$.  These generators reproduce the same degeneracy structure as (\ref{degeneracys}). Such operators were also found for the 1D multispecies CSM~\cite{MMS03}, a different physical model.  The multispecies model generalizes the full CSM by allowing the masses $m_i$ and interactions between particles $\lambda_{ij}$ to vary, where $\forall \lambda_{ij}>\frac{1}{2}$.  By absorbing the truncated range into the interaction strength so that $\lambda\rightarrow\lambda_{ij}=\lambda\theta(r-|i-j|)$, $\forall i,j$  and taking $m_i=m$, $\forall i$, the mathematical construction of the multispecies model is related to the TCSM.  However, the essential difference between the two models is that the TCSM describes a screened interaction between particles, whereas the multispecies model describes distinguishable particles with full range interactions of varying strength.  In particular, the wavefunction of the multispecies CSM  has compact support on a given sector, as the particles are distinguishable. Different sectors correspond to different physical realizations of the ordering of the particles, that are preserved under the time evolution generated by the system Hamiltonian.  By contrast,  particles in the symmetrized TCSM  are indistinguishable and different sectors are explored as a result of scattering among particles. As a result, the behavior of non-local correlation functions such as the one-body reduced density matrix or the momentum distribution differs in these two models. 


For $r=\N-1$, the number of constraints on $P_k(\mathbf{x})$ decreases to $\mathcal{N}_c=M(k-2)$.   Given $k$, there are $M(k)-M(k-2)$ unique solutions for $P_k(\mathbf{x})$. For example, table~\ref{Coefficients2} shows the constraints for $3\leq k \leq 5$.

The corresponding level degeneracy for each $s=2n+k$ is $M(s)$, as it satisfies
\begin{eqnarray}
\sum_{n,k}\Big[M(k)-M(k-2)\Big]\delta(2n+k-s)=M(s),
\end{eqnarray}
\noindent 
and the excited states, $L^{\nu}_n(\tilde{\omega} \rho^2)P_k(\mathbf{x})$, can be expressed as a linear combination of the Hi-Jack Polynomials~\cite{ujino1996algebraic}.  The level degeneracy directly relates to Calogero's result~\cite{Calogero71}, where the $\N$ quantum numbers $\{n_l\}$ are solutions to
\begin{eqnarray}
s=\sum^\N_{l=1} l n_l,
\end{eqnarray} 
\noindent
which holds for all $k$.  Thus, solving for $P_k(\mathbf{x})$ using a linear combination of monomials reproduces the full spectrum of the CSM in the limit $r=\N-1$.

One can explicitly calculate the excited states using the constraints provided in table~\ref{Coefficients} for $k\leq 5$ and $\N\leq K$ .  Figure~\ref{fig:N5ralldensity} shows the density profile of the first three excited states for $\N=4$ particles with fixed interaction strength $\lambda=1$ and varying interaction range $r\leq 3$.  While the coefficients vary with $r$, $\lambda$ and $N$, the behavior of the density $n_{n,k}(x)$ is largely dominated by the ground state density $n(x)$.   Specifically, increasing the interaction range $r$  broadens the density profile $n(x)$, see Fig.~\ref{fig:N5ralldensity}, as observed as well in the ground state case.   In addition, the local maxima of the density become more apparent, signaling a higher degree of spatial antibunching.


\section{Conclusions and Outlook}

We have introduced a family of models describing  one-dimensional particles confined in  a harmonic potential and subject to  inverse-square interactions  among a finite number of neighbors.  
This family of  truncated Calogero-Sutherland models involves pairwise interactions that can be repulsive or attractive and a  three-body contribution that is always  attractive. It interpolates between the  Calogero-Sutherland model with full range interactions, when the three-body term vanishes, and the model introduced by Jain and Khare.
The system can be understood as a quantum chain in the continuum space or a quantum fluid after restoring full permutation symmetry. In the latter case, it includes the Tonks-Girardeau gas that describes impenetrable bosons in one spatial dimension as well as a novel  extension with truncated interactions.
The TCSM remains quasi-exactly solvable in spite of the tunable strength and range of the interactions. In addition, we have found a particular set of collective excitations in terms of symmetric polynomials.  By numerically computing the density profile and the one-body reduced density matrix we have demonstrated that increasing the interaction strength and range leads to spatial antibunching and suppresses off-diagonal long-range order.  The tail of the one-body reduced density matrix decays according to an $r$-dependent exponent, which increases with increasing interaction range $r$.  This decay manisfests in the the  momentum distribution that exhibits a central peak at $k=0$ and broadens with increasing $r$.  

This family of truncated models can be extended in a wide variety of ways to account for particles with internal structure \cite{Kawakami93,VOK94}, multiple species \cite{MMS03}, higher dimensions \cite{MMS04}, anyonic and fermionic exchange statistics, and modified interactions and confining potentials \cite{Sutherland04}. 
One may envision as well an extension of our model to a variety of root systems with non-traditional reflection symmetries \cite{Turbiner11}.


{\it Acknowlegments.---} 
This work is supported by UMass Boston (project P20150000029279). MO further acknowledges the support from the US National Science Foundation (PHY-1402249) and the US Office of Naval Research (N00014-12-1-0400).

\appendix*

\section{}
\label{Appendix}
We start by assuming that the wavefunction has a separable form $\Psi=\phi_n(\mathbf{x})\varphi(\mathbf{x})$ and make the similarity transformation,
\begin{eqnarray}
\tilde{\mathcal{H}}&=&\varphi^{-1}(\mathbf{x})\mathcal{H}\varphi(\mathbf{x}) \nonumber \\
& =& \sum^N_{i=1}\Big(-\frac{\hbar^2}{2m}\frac{\partial^2}{\partial^2 {x_i}} +\frac{1}{2}m\omega^2 x^2_i\Big) \nonumber \\
&  -& \frac{\hbar^2 \lambda}{ m}\sum_{\substack{i<j,|i-j|\leq r}}\frac{1}{(x_i - x_j)}\Big[\frac{\partial}{\partial {x_i}}-\frac{\partial}{\partial {x_j}}\Big]
\label{reduced} \end{eqnarray}
\noindent

We then separate the center of mass and relative degrees of freedom of the reduced Hamiltonian (Eqn.~\ref{reduced}),
\begin{eqnarray}
\tilde{\mathcal{H}}&=&\tilde{\mathcal{H}}_{cm}+\tilde{\mathcal{H}}_{rel} \label{newH} \\
\tilde{\mathcal{H}}_{cm} &=&-\frac{1}{2m\N}\frac{\partial^2}{\partial^2 X}+\frac{\N M\omega^2}{2}X^2 \label{COMH} \\ 
\tilde{\mathcal{H}}_{rel}&=& \frac{\hbar^2}{2m}\sum^{\N}_{i=1} -\frac{\partial^2}{\partial \zeta_i^2}+\tilde{\omega}^2 \zeta^2_i \nonumber \\
& &\ \ \ \  - \frac{\hbar^2\lambda}{m}\sum_{\substack{i<j \\ |i-j|\leq r}}\frac{1}{(\zeta_i-\zeta_j)}\Big[\frac{\partial}{\partial \zeta_i}-\frac{\partial}{\partial {\zeta_j}}\Big] \label{relativeH}
\end{eqnarray}
\noindent
where $X=\frac{1}{\N}\sum^\N_{i=1}x_i$ is the center of mass and $\frac{\partial}{\partial_X} = \sum^\N_i \frac{\partial}{\partial_{x_i}}$. The relative variables $\zeta_i=x_i-X$ are linearly dependent, but can be rewritten in terms of $\N-1$ linearly independent variables.  

The center of mass Hamiltonian (Eqn.~\ref{COMH}) is simply that of a single harmonic oscillator, with creation and annihilation operators
\begin{eqnarray}
a^{\pm}_X &=& \frac{1}{\sqrt{2}}\Big(\sqrt{\N\tilde{\omega}}X\mp \frac{1}{\sqrt{\N\tilde{\omega}}}\frac{\partial}{\partial X} \Big) \\
\tilde{\mathcal{H}}_{CM} &=& \frac{\hbar\omega}{2}\{a^{-}_X,a^{+}_X\} =\hbar\omega\Big(\hat{N}_X+\frac{1}{2}\Big)
\end{eqnarray}
 \begin{eqnarray}
[\tilde{\mathcal{H}}_{CM}, a^{\pm}_X ]  =  \pm \hbar \omega a^{\pm}_X  
\end{eqnarray}
The relative Hamiltonian (Eqn.~\ref{relativeH}) can be rewritten in terms of previously defined SU(1,1) generators  
\begin{eqnarray}
\tilde{\mathcal{H}}_{rel}=m \omega^2 D^{'}_{-}-\frac{\hbar^2}{m}D^{'}_{+} \label{relativeHDs} 
\end{eqnarray}
\noindent
where the center of mass has now been factored out.  The ground state energy that formerly appeared in $K'$ is now $E^{0,rel}_{\N,\lambda,r}=\frac{\hbar\omega}{2}\Big[\N-1+\lambda r(2\N-r-1)\Big]$, the ground state energy of the $N-1$ linearly independent relative degrees of freedom.
For $\tilde{\mathcal{H}_{rel}}$ we can define a pair of creation and annihilation operators~\cite{MMS04},
\begin{eqnarray}
a^{\pm}_{rel}&=&\frac{1}{2}\Big(\frac{D'_{+}}{\tilde{\omega}}+\tilde{ \omega} D'_{-} \Big)\pm K' \label{creationannihilationHrel}
\end{eqnarray}
The commutation of these operators yields the relative Hamiltonian,
\begin{eqnarray}
\tilde{\mathcal{H}}_{rel}= \hbar \omega & &\Big[a^{-}_{rel},a^{+}_{rel}\Big]  = \hbar\omega \Big(2\hat{N}_{rel} +\frac{E^{0,rel}_{\N,\lambda,r}}{\hbar \omega}\Big) \label{Hrellad}\\
& & \Big[\tilde{\mathcal{H}}_{rel},a^{\pm}_{rel}\Big]=\pm 2\hbar \omega a^{\pm}_{rel} \label{numop}
\end{eqnarray}
\noindent
where the number operator is defined as $\hat{N}_{rel}=\frac{\tilde{\mathcal{H}}_{rel}}{2}-E^{0,rel}_{\N,\lambda,r}$.  The commutation relation between the pair of operators has a general non-canonical form,
\begin{eqnarray}
a^{-}a^{+}-a^{+}a^{-}=F(\hat{N})
\end{eqnarray}
\noindent
where $F(\hat{N})$ is a function of the number operator.  Therefore, the pair of creation and annihilation operators that give (\ref{Hrellad}) characterize a deformed single-mode oscillator~\cite{meljanac1994unified,meljanac1996unified}.  The relationship between $a^{\pm}_{rel}$ and undeformed bosonic creation and annihilation operators $\{b_{rel},b^{+}_{rel}\}$ is,
\begin{eqnarray}
a^{+}_{rel}=\sqrt{\frac{\phi(\hat{N}_{rel})}{\hat{N}_{rel}}}b^{+}_{rel} =\sqrt{\hat{N}_{rel}-1+\frac{E^{0,rel}_{\N,\lambda,r}}{\hbar\omega}}b^{+}_{rel}\label{bosonoperatorp} \\
a^{-}_{rel}=b_{rel}\sqrt{\frac{\phi(\hat{N}_{rel})}{\hat{N}_{rel}}} = b_{rel}\sqrt{\hat{N}_{rel}-1+\frac{E^{0,rel}_{\N,\lambda,r}}{\hbar\omega}} \label{bosonoperatorm}
\end{eqnarray}
where $\phi(\hat{N}_{rel})$ is a function of the number operator,
\begin{eqnarray}
\phi(\hat{N}_{rel})=\hat{N}_{rel}\Big(\hat{N}_{rel}-1+\frac{E^{0,rel}_{N,r,\lambda}}{\hbar\omega}\Big)
\end{eqnarray}
The excited states of (\ref{newH}) are given by
\begin{eqnarray}
|n_X\ \ n_r\rangle = \frac{(a^{+}_X)^{n_X}}{\sqrt{n_X!}}\frac{(a^{+}_r)^{n_r}}{\sqrt{\phi(n_r)!}}|0\rangle
\end{eqnarray} 
\noindent
which act on the vaccuum state,
\begin{eqnarray}
\langle X\ \mathbf{\zeta}|0 \rangle=\exp\Big[-\frac{\N\tilde{\omega}}{2}X^2\Big]\exp\Big[-\frac{\tilde{\omega}}{2}\sum^{\N}_{i=1} \zeta^2_i \Big]
\end{eqnarray} 
\noindent
The total energy is linear in the quantum numbers $n_r$ and $n_X$,
\begin{eqnarray}
E_{n_r,n_X}=\hbar \omega (n_X+2n_r)=\hbar \omega s
\end{eqnarray}

We then construct generators of the $\mathfrak{sl}(2,\mathbb{C})$ algebra using the creation and annihilation operators of the two uncoupled oscillators $\{a^{\pm}_X,b^{\pm}_r\}$ ~\cite{MMS04,jonke2002bosonic},
\begin{eqnarray}
J^{+}&=& (a^{+}_{X})^2b_{rel}\frac{1}{\sqrt{2\Big(\hat{N}_X+1\Big)}} \\
J^{-}&=&b^{+}_{rel}(a^{-}_{X})^2\frac{1}{\sqrt{2\Big(\hat{N}_X-1\Big)}} \\
J_z &=&\frac{1}{4}\Big(\hat{N}_X-2\hat{N}_{rel}\Big) 
\end{eqnarray}
\noindent
which act on states $|n_X \ n_r\rangle$ for $s=n_X + 2n_r \geq 2$.  
 The Casimir operator reads,
\begin{eqnarray}
J^2_0&=& J^2_z+\frac{1}{2}\Big[ J_{+} J_{-}+J_{-} J_{+}\Big] 
\end{eqnarray}
with eigenvalue
\begin{eqnarray}
J^2_0|n_X\ \ n_r\rangle=j(j+1)|n_X\ \ n_r\rangle
\end{eqnarray}
\noindent
where $j=\frac{1}{4}(n_X+2n_r)=\frac{s}{4}$.  Each irreducible representation is given by $j=\frac{1}{2},\frac{3}{4},1,....$, with dimension
\begin{eqnarray}
d(j) = [2j]+1= \Big[\frac{s}{2}\Big]+1
\end{eqnarray}
\noindent
where $[x]$ outputs the floor of $x$.



 \bibliography{BiblioTCS}

\end{document}